\documentclass[12pt,showpacs,preprintnumbers,floatfix]{article}
\pdfoutput=1
\usepackage{graphicx} 
\usepackage{dcolumn} 
\usepackage{bm} 
\usepackage{amsfonts}
\usepackage{amsthm}
\usepackage{amsmath}
\usepackage{amssymb}
\usepackage{epsfig}
\usepackage{color}
\usepackage{textcomp}
\usepackage{hyperref}
\usepackage{titlesec}
\usepackage{subfigure}
\usepackage[utf8]{inputenc}
\usepackage[english]{babel}
\usepackage[usenames,dvipsnames]{xcolor}

\def\be{\begin{equation}}
\def\ee{\end{equation}}
\def\ba{\begin{eqnarray}}
\def\ea{\end{eqnarray}}

\def\bdm{\begin{displaymath}}
\def\edm{\end{displaymath}}

\def\bq{\begin{quote}}
\def\eq{\end{quote}}

 at 10truept

\newcommand{\bea}{\begin{eqnarray}}
\newcommand{\eea}{\end{eqnarray}}

\newcommand{\bi}{\begin{itemize}}
\newcommand{\ei}{\end{itemize}}

\newcommand{\beq}{\begin{equation}}
\newcommand{\eeq}{\end{equation}}
\newcommand{\beqa}{\begin{eqnarray}}
\newcommand{\eeqa}{\end{eqnarray}}

\def\ltap{\ \raise.3ex\hbox{$<$\kern-.75em\lower1ex\hbox{$\sim$}}\ }
\def\gtap{\ \raise.3ex\hbox{$>$\kern-.75em\lower1ex\hbox{$\sim$}}\ }
\def\gl{\ \raise.5ex\hbox{$>$}\kern-.8em\lower.5ex\hbox{$<$}\ }
\def\roughly#1{\raise.3ex\hbox{$#1$\kern-.75em\lower1ex\hbox{$\sim$}}}

\begin{document}

\thispagestyle{empty}
\begin{flushright}
July 2017
\end{flushright}
\vspace*{1.7cm}
\begin{center}
{\Large \bf Anthropics of Aluminum-26 Decay\vspace{.2cm}\\ and Biological Homochirality}\\

\vspace*{1.2cm} {\large McCullen Sandora\footnote{\tt
mccullen.sandora@tufts.edu}}\\
\vspace{.5cm}
{\it  $^{}$Institute of Cosmology, Department of Physics and Astronomy}\\
	{\it  Tufts University, Medford, MA 02155, USA}

\vspace{2cm} ABSTRACT
\end{center}
Results of recent experiment reinstate feasibility to the hypothesis that biomolecular homochirality originates from beta decay.  Coupled with hints that this process occurred extraterrestrially suggests aluminum-26 as the most likely source.  If true, then its appropriateness is highly dependent on the half-life and energy of this decay.  Demanding that this mechanism hold places new constraints on the anthropically allowed range for multiple parameters, including the electron mass, difference between up and down quark masses, the fine structure constant, and the electroweak scale.  These new constraints on particle masses are tighter than those previously found.  However, one edge of the allowed region is nearly degenerate with an existing bound, which, using what is termed here as `the principle of noncoincident peril', is argued to be a strong indicator that the fine structure constant must be an environmental parameter in the multiverse.

\vfill \setcounter{page}{0} \setcounter{footnote}{0}
\newpage

\section{Introduction}
Does our universe possess any special properties that make it more conducive to life, on the molecular level, compared to  generic universes?  This is not a new question: in fact, some of the earliest anthropic arguments were in regards to biochemistry.   In the book {\it The Fitness of the Environment\,} written early last century, Henderson \cite{fitness} detailed the myriad ways in which both water and carbonic acid (carbon dioxide dissolved in water) seem specially suited for life to develop.  He used these arguments as evidence for preternatural selection of some microscopic characteristics of our world, which render it unusually well suited for our existence.

Today many of these arguments have fallen out of favor, for several reasons. The first of which is that we now know that many of the properties of chemicals are dictated by basic orbital structure set by the charges of the constituent atoms, and not by any  conceivable fundamental parameters that can be altered.   In this way, the interactions between atoms that form the basis for chemistry are somewhat rigid, and their basic properties cannot be  varied.  

While these gross features likely remain unchanged in other possible universes, some of the finer details can in fact depend on fundamental parameters, such as the proton mass, neutron mass, electron mass, and the fine structure constant.   However, it is usually found that the dependence of chemical properties on these parameters is relatively weak when compared to other aspects of the universe.   For instance, ab initio calculations of water molecules find that in order to induce a 10$\%$ shift in the dipole nature of water, the electron mass must be varied by a factor of 20, and the fine structure constant must be varied by factor of 7 \cite{water}. These are rather extreme deviations for these fundamental constants, and changes in other aspects of the world such as nuclear and stellar structure are affected far before these thresholds are reached. 

This relative insensitivity is expected to be a generic feature of the chemical world for one simple reason: the chemical binding energies represent intermolecular forces which are a few orders of magnitude lower than electronic transitions, and a few orders of magnitude lower still than nuclear energies. Any change to chemical energies would be proportionately smaller by these same factors, and the effects therefore weaker.  Thus, it would seem that for the most part, any universe that manages to give rise to the diversity of stable atoms such as ours would automatically fulfill the conditions for a complex chemistry to operate.

 The second reason Henderson's argument has of fallen out of favor is that we now know that, unlike the laws of physics, which, to the best of our knowledge, seem to be rigid, and uniform throughout the entire universe, the laws of chemistry are heavily dependent on environment. The rate equations are dependent on a number of variables, including temperature, pressure, and concentrations of constituents present, and so the operationally dominant chemical reactions are dependent on the environment. Indeed, it has been argued \cite{hard} that this flexibility is essential for the emergence of such complexity to be worthy the of the term life: this stems from the observation that the likelihood of any given state in a system as complex as the one we inhabit is unforeseeably sensitive not only to the initial conditions, but to the laws which subsequently govern their evolution.  
 
 At the time of Henderson's writing, people were under the impression that both Venus and Mars were relatively Earthlike in temperature and atmosphere.  Subsequent investigations of planets in our solar system (and now elsewhere) have revealed a diversity far greater than imagined, with environments giving rise to complex chemical reactions much different from those found on Earth. Thus, one may suspect that even in a universe whose chemistry is substantially altered from ours, it would contain at least some environment capable of yielding comparable complexity to that of the Earth. Our habitual lack of imagination, along with the continual surprising discoveries of the diversity found within our universe, should make one wary of dismissing other universes as being sterile.

Thus, our particular biochemistry seems to be just one of a great number of possibilities that was particularly good at replicating and information storage early in the history of Earth (or possibly before). Is there any feature it possesses that we may suspect is universal? Indeed there is: it has been known for about a century and a half that on a biomolecular level, life is chiral. That is, when given a choice between two different configurations of its constituent molecules, it chooses to utilize one uniformly over the other. 

The origin of this scenario remains a mystery to this day. The idea that this chirality was inherited by the laws of physics was proposed by Vester and Ulbricht almost immediately after it was found that the weak nuclear force violates parity at a fundamental level \cite{VU}. However, any direct influence on chemistry is extraordinarily weak, making a connection between the two rather tenuous. In light of this, a scenario has been proposed by Cline \cite{cline} for an indirect connection whereby chemistry would inherit a chiral imprint from the decay of certain radioactive isotopes, namely aluminum-26. If this scenario is true, then the radioactive properties of aluminum-26, such as its lifetime and energy, are crucial for the basic requirements for life. These, in contrast to some of the grosser properties of chemistry, are readily altered by minute changes in the laws of physics. There are several aspects of this isotope that make it unusually suited for the task of imprinting handedness in a chirally pure medium, as we will detail below. Though this scenario remains speculative, it may soon become testable with current technology. If such a mechanism is found to be the dominant source of chirality in our universe, then it would give an indication that the laws of physics have conspired in this way to make biochemistry more feasible.

In section \ref{lias} we review the current status of biomolecular homochirality, and give the context and evidence for which aluminum-26 may play a role.  In section \ref{nuphalde} we track the nuclear properties of aluminum-26 in terms of fundamental parameters.  In section \ref{pade} we interpret the dependences on these parameters, and compare them to preexisting bounds.  In section \ref{codi} we conclude.
 
\section{Life's Asymmetry}\label{lias}
The origin of biomolecular homochirality has been one of the mysteries of life for over a century, when Louis Pasteur noticed that the polarization of light rotated when passing through some biological liquids.  It was deduced that this arises because many chemical compounds have two possible configurations, dubbed enantiomers, and life makes use of only one of these (for reviews see \cite{podlech,gk,bonner,keszthelyi}).  In fact, of all the amino acids, all but the simplest one, glycine, come in two varieties, and, (practically) without fail, life only uses the left handed version\footnote{This handedness arises because all these molecules are based off of carbon, which forms four bonds.  The first participates in linking to others to form a protein chain, and the remaining three may be arranged by size either clockwise or counterclockwise around this axis.}.  Similarly, many sugars, including those used in RNA and DNA, only contribute in their right handed configuration.

It is important to note that this chirality is necessary for life as we understand it.  Firstly, experiments that have made DNA strands containing chiral defects have shown that they cannot form their pair bonds properly, which would have prevented its ubiquitous use as information storage\cite{gk}.  It is likely that even if a chirally mixed strand could be used adequately for single celled organisms, the mutation rates would be too high for long lived, multi-celled organisms to survive\cite{cline}.  Secondly, it is possible to estimate the length that replicating strands of proto-RNA can attain in a mixture of a given chirality, as done in \cite{gk}.  There they report that the limiting size of these replicators in a racemic (completely mixed) medium is much too low to contain the information needed for self-replication.  All this suggests that the prebiotic medium itself started in at least a partially chiral state, rather than the alternative that biological processes drove the system to be chirally pure through some selection process.\\

\noindent{\bf Leading Hypotheses:}
This leaves the question of what processes could have driven the prebiotic system at least partially towards chirality, and the answers put forward, though spanning a great number of academic disciplines, fall broadly into three categories\cite{bonner}: (i) that some regional process is responsible, (ii) that a spontaneous amplification of a stochastic initial fluctuation occurred, or (iii) that it is inherited through the parity asymmetry of fundamental physics.  Each of these scenarios has its challenges.  For the first, many processes, including the Earth's magnetic field, circularly polarized light from the sun, adsorption onto clays or quartz, and even the Coriolis force have been invoked over the years, but the efficacy of each has been hard to convincingly demonstrate\cite{keszthelyi}.  

The spontaneous symmetry breaking scenario aims to find an environment where the chemical reactions naturally drive the system to select one parity over the other, even in the absence of an innate preference (see e.g. \cite{morozov}).  Abstractly (in terms of mathematical chemistry modeling), the conditions for this to occur can be succinctly expressed as a tendency for the racemic state to be unstable.  For any given chemical reaction network, it is straightforward to determine whether or not this is the case.  Most chemical reactions do not exhibit this behavior, and those that do will only work if some environmental condition specific to the exact scenario is satisfied, so this mechanism is not guaranteed to occur.  Existing proposals of systems that can accomplish are not without their difficulties: simple autocatalysis is not sufficient, so in \cite{morozov} superautocatalytic processes were appealed to.  Preferential polymerization seems a likely scenario, but it was shown in \cite{agk} that unless polymerization selected for chiral purity with near perfect fidelity, the mechanism would be  inoperational.  The challenge, then, is to find a situation where plausibly small molecules are able to encode for such stringent construction mechanisms.  A recent proposal \cite{jbg} relies on inherent noise of small aggregates, but so far has only been demonstrated to be successful with less than $10^3$ molecules.  Of course, there are significant unknowns accompanying any origin of life scenario, so one of these purely chemical reactions may indeed turn out to be incorporated in a natural way, but this scenario is far from empirically established.



The third option relies on the inherent parity breaking of the weak nuclear force.  This can be manifested several ways: the first is directly, through the energy level splitting of the enantiomers via weak neutral currents.  The energy difference here is exceedingly tiny compared to typical thermal energies, usually around $10^{-17}$\cite{pdiff}, meaning that for this process to be operational, a large amount of reactants must participate in the origin scenario, in order to not be swamped by random noise.  Alternatively, the asymmetry may have been built up indirectly through some beta decay process, which preferentially emits either an electron or positron of definite handedness.  Neither of these methods would result in perfect symmetry breaking, but they are potentially able to imprint a slight preference in an initially racemic mixture, that may then be subjected to a secondary stage of amplification in order to become a fully chiral medium\cite{lwc}.  This initial asymmetry may have been crucial, as chemical reactions that are able to amplify an initial signal are quite commonplace \cite{nonlin}, but they do require a seed to operate.

The main problem with any scenario, aside from the initial preparation of an environment that can yield an effect, is that any chirality that is present tends to racemize with a half life on the order of $10^4-10^6$ years, depending on the chemical species, which is much shorter than the timescales involved in many of the processes required to build up a significant fraction of one parity \cite{gk}.  This racemization is governed by a quantum tunneling between two practically degenerate states, and so is given by $t\sim \tau e^{\sqrt{2mV_0}\Delta x}$, where $\tau$ is the chemical interaction time, and is exponentially suppressed by the tunneling under the potential barrier of height $V_0$ and length $\Delta x$.  Because the exponent is fixed, this provides a clue to which type of environment would be able to retain a significant chirality: one that is cold and extremely sparse.  This suggests that chirality would have a much easier time building up in space than it would on the surface of a planet \cite{bonner}.  There, chirality can be stabilized over long periods of time, but the problem with this scenario is that many of the usually suspected processes take much longer in this environment as well, so that chirality may have a hard time developing in the first place\cite{gk}.\\

\noindent{\bf Chirality in Space:}
There is in fact evidence that extraterrestrial molecules already contain a preference for a particular chirality.  This comes from the Murchison meteorite \cite{murch}, which fell in 1969, where it was found that several amino acids carry a definite handedness.  Suspicion that this may be contamination is diminished by the observation that this asymmetry is even present for amino acids that are not used by life.  Still, it is known that the shock of a meteorite entering the atmosphere can produce amino acids \cite{cs,shocks}, so that even though no way of preferentially producing one chirality in this process has been proposed, the observation does not conclusively indicate an extraterrestrial origin.  Subsequent experiments have shown that prebiotic molecules are in fact present in the solar system: the STARDUST instrument collected and returned amino acids from a comet\cite{stardust}, and the COSAC instrument aboard PHILAE detected several when it landed on 67/P Churyumov-Gerasimenko \cite{cosac}.  (Unfortunately, both the experiments undertaken to measure extraterrestrial chirality directly in the past few years, COSAC aboard PHILAE and MOMA aboard EXOMARS, resulted in malfunctions that prevented the experiments from being carried through to completion.) No processes capable of inducing chirality on the surfaces of comets and asteroids are known, though, and \cite{candc} concluded that any chiral preference must have been established in these molecules before they reached the surface.  

Indeed, recent observations now indicate that biological and prebiological molecules are present in molecular clouds and in protostellar nebulae.  In \cite{glue}, the chiral molecule propylene oxide, though smaller than an amino acid, was discovered in interstellar space.  This holds promise that organic chemistry is a ubiquitous process in the galaxy, and that chirality may be imprinted at a very early stage in this process.  No currently planned experiments are capable of measuring the chirality of these molecules, but it should be possible in principle, without any substantial improvement on current technology.

Several processes have been proposed that could lead to a buildup of chirality in molecular clouds.  The first is the presence of circularly polarized light, which has been observed \cite{cplexpt} in star forming regions.  Though it has been determined that the level of polarized radiation coming from supernova or stellar winds is incapable of inducing the observed level of chirality, Mie scattering of unpolarized light by aligned dust grains is capable of imprinting a preference in the molecular cloud \cite{CPLtheory}.  The trouble with this scenario is that even though each individual supernova may produce light of a definite polarization, depending on its geometry with respect to the molecular cloud, aggregating the effect over many supernovae would tend to wash this effect out \cite{cline}.  

However, there is a mechanism by which supernovae may imprint uniform chirality on the medium: radioactive beta decays.  Supernovae produce an abundance of radioactive elements, and because of the inherent handedness of the weak force, when they decay they emit electrons with left handed helicity (and positrons with right handed helicity)\footnote{For completeness we also mention a scenario where the accompanying neutrinos imprint chirality on the medium, though the cross sections are orders of magnitude weaker \cite{cline}.}.  The majority of the time, the energy exchanged is of the order a few electron-volts (eV), and so will just excite a rotational state of the molecule.  A small fraction of the interactions, however, will destroy the parent molecule.  The structure of the prebiotic molecule designates a preferred helicity for the molecular electrons, which was shown in \cite{pdiff} to be the right sign to explain both the predominance of L-amino acids and D-sugars.

The main drawback of this scenario is that is has been historically extremely difficult to demonstrate that the level of asymmetry necessary is attained.  This long struggle begins with the null results of Vester and Ulbricht \cite{VU}, and goes through decades of controversy and irreproducible claims (for a brief review see \cite{dg,gk}).  However, this criticism may finally be put to rest: in a recent article \cite{dg}(designated a featured article and editor's suggestion in Physical Review Letters), it was experimentally established that polarized electrons incident on a gas are capable of imprinting a small degree of chirality, of the order of $\sim3\times10^{-4}$.  Though tiny, this is a sufficient amount to be subsequently amplified by a second stage of chemical evolution.  The molecules used were not amino acids, but this demonstration removes the main source of doubt from this hypothesis.

The exact physical mechanism responsible for this remains open to interpretation.  One possibility, as advocated by \cite{dg}, is that the molecule temporarily captures the incident electron, exciting and possibly destroying the molecule.  Another possibility is that the incident particle scatters off an outer shell electron as it streams through the interstellar medium.  To leading order, this is independent of chirality of the electrons, but cross terms depend on their relative helicities \cite{meiring}.  In order for this scattering to conserve angular momentum, some must be exchanged with the parent molecule, which some fraction of the time result in its destruction.  Key to this new result is that the incident electron energies were of order a few eV, leading to significantly enhanced cross sections for this process to be operational.  It is important to note, then, that the initially relativistic electrons (and especially positrons) are capable of retaining their helicities as they lose energy in the interstellar medium.

The question of which isotope(s) may have been responsible remains.  Of all the candidates, only aluminum-26 is produced in enough abundance, has a half-life relevant for the timescales of the dynamics ($717,000$ years), and decays through the channels necessary to communicate the parity asymmetry of particle physics to the chemical realm.  This will be discussed further in section \ref{nuphalde}.

In \cite{cline} it was demonstrated that this process is capable of inducing a sufficient level of chirality in the medium.  For typical values of molecular cloud density, a given decay product interacts with $10^5$ different electrons before it becomes nonrelativistic, and for reasonable concentrations produced by supernovae injection is much more than sufficient to polarize the entire medium, by many orders of magnitude.  Though there are several senses in which this estimate is optimistic, more conservative values still yield well over the required values.

It is worth reiterating the main difference between the two viable interstellar mechanisms: circularly polarized light, if the main source of chirality, will leave a random chirality in each molecular cloud, whereas the decay of radioactive nuclei would leave the same chirality in each cloud.  Since there is no obvious obstruction to measuring the chirality of molecular clouds even with existing technology \cite{glue,patty}, we may soon be able to definitively favor which of these scenarios is responsible.

\section{The Nuclear Physics of Aluminum Decay}\label{nuphalde}

We now turn to the microphysical basis of the properties of aluminum-26 that make it the ideal candidate for biomolecular chirality.  The nucleus of ${}^{26}$Al is in a $5^+$ spin state.  It predominantly decays through positron emission to an excited state of magnesium-26 with spin $2^+$\cite{CEA}.  Such a large exchange of angular momentum makes this a 2nd forbidden transition, meaning that it is mediated by matrix elements corresponding to second order terms in the expansion of the electron wave function\cite{laubitz}.  This is partially what is responsible for the long lifetime of this decay:  typical 2nd forbidden lifetimes fall within the range $10^4-10^6$ years\cite{krane}.  Even for this type of transition, however, aluminum-26 has one of the longest lifetimes.  The reason for this is because the energy difference between the two states is very close to the electron capture threshold $m_e$.  Below this threshold, positron emission becomes energetically impossible, and the decay can only occur through electron capture.  As it stands, this decay occurs through electron capture 15$\%$ of the time.  If the energy difference were much lower, electron capture would be the dominant channel, which would emit a noninteracting neutrino and randomly polarized photon.  Much higher, and the half life of the decay would be much shorter \cite{br}, like the vast majority of $\beta^+$ decays.  Either of these would prevent the generation of molecular chirality, and so would lead to a less fecund universe.\\

\noindent{\bf Half Life:}
The half-life for beta decay is \cite{krane}
\beq
t_{1/2}=\frac{2\pi^3\ln2\,v^4}{m_e^5 f(w_0)}
\eeq
Where $v$ is the Higgs vacuum expectation value (VEV), related to Fermi's constant by $v^2=\sqrt{2}G_F$, $m_e$ is the electron mass, and $f(w_0)$ is the Fermi integral.  This takes into account the specific interaction that mediates this decay, and the maximum energy the electron can have, $w_0$, made dimensionless by dividing by the electron mass.  This interaction is found to be induced by the $S_{ijk}$ matrix element of \cite{ku,laubitz}, which involves the insertion of two position and one spin operator.  For our purposes we use the small charge limit, $\alpha Z\ll1$.  This does not introduce significant errors \cite{ku}, and it is not possible to distinguish this approximation from the true curve by eye.  The Fermi integral is then found to be
\beq
f(w_0)=f_0\int_1^{w_0}dw \sqrt{w^2-1}\,w\,(w_0-w)^2\frac{(-3 + 4 w^2 - 2 w w_0 + w_0^2) (-1 + 4 w^2 - 6 w w_0 + 3 w_0^2)}{3240}
\eeq
The coefficient $f_0$ is an inherent ambiguity of the formalism, representing the choice of where in the nucleus the Coulomb potential is to be evaluated \cite{ku}.  We set this parameter $f_0=1/32.7$ by enforcing the half life to correspond to the observed value when the constants take their measured values.  Then the integral can be done analytically to yield
\bea
f(w_0)=\frac{f_0}{4,082,400}\Bigg[\sqrt{w_0^2-1}\left(8-536w_0^2-3,417w_0^4-515w_0^6+50w_0^8\right)\\+315w_0^3(5+9w_0^2)\log\left(w_0+\sqrt{w_0^2-1}\right)\Bigg]
\eea
This has the limiting behavior
\begin{equation}
f(w_0)\rightarrow \left\{
 \begin{array}{cl}
  \frac{f_0}{374,220}(w_0^2-1)^{11/2} &\quad w_0\rightarrow 1\\
   \frac{f_0}{81,648}w_0^9 & \quad w_0\rightarrow \infty
 \end{array} \right.
\end{equation}
Though the observed value of $w_0=2.3$ \cite{CEA} is too intermediate for this function to be well described by either of these approximations, and so the full expression is required.  Notice that the dependence on energy is so strong in either case that it reverses the half life's dependence on the electron mass: because the phase space factors strongly dictate the rate, an increase in mass will increase the half life.  This is typical for such forbidden decay channels.\\

\noindent{\bf Energy:}
We now need to model how the energy released depends on physical parameters.  We base this calculation off the semi-empirical mass formula (SEMF), which does a fair job in determining the binding energies of nuclei (usually around 10$\%$ accuracy) based off a simple phenomenological prescription.  This takes into account strong interactions between nearest neighbor nuclei, a surface energy term, the Coulomb interaction among protons, Fermi repulsion, and pair bonding.  These energy contributions are given by five phenomenological parameters that are set as a best fit for the entire range of observed nuclei.  Up to contributions that are irrelevant to beta decay, the binding energy is
\beq
E_b=\left(-.72 \frac{\alpha}{\alpha_0}\frac{Z(Z-1)}{A^{1/3}}-23\frac{(A-2Z)^2}{A}-12\frac{(-1)^Z}{A^{1/2}}\right)\frac{m_p}{m_{p0}}\text{MeV}
\eeq
Note that the entire expression scales linearly with the proton mass, as it is the natural energy scale in the problem.  Binding energies in the SEMF framework do not make a distinction between the mass of the proton and neutron, though the total mass of the nucleus accounts for this difference.

Using this, we find that the binding energy of aluminum-26 is $E_b({}^{26}\text{Al})=209\text{MeV}$ for our values of proton mass and fine structure constant, in good agreement with the experimental value $E_b({}^{26}\text{Al})=211.9\text{MeV}$.  The binding energy of magnesium-26 is a bit low, $E_b({}^{26}\text{Mg})=212\text{MeV}$, as compared with the experimental value $E_b({}^{26}\text{Mg})=216.7\text{MeV}$.  While this agreement is as good as one can hope for with this simplistic model, the SEMF value would not provide sufficient energy for this beta decay process to occur.  Therefore, we recalibrate the difference by subtracting a fiducial energy $E_{\text{shift}}=6.52(m_p/m_{p0})\text{MeV}$, in order to yield the correct lifetime, while capturing dependences of energies on fundamental parameters as realistically as possible.  A full numerical QCD calculation would be desirable, to find the dependence on underlying parameters more precisely, though one has never been performed for such a heavy nucleus.  Even at the quantitative level, our estimates will not be too far from their exact values, though they will be subject to slight shifts.

We must also note that the transition is not between the ground states of these nuclei, but between the ground state of aluminum and an excited state of magnesium, with energy $E_2=1.8\text{MeV}$.  This is a rotational excited state, and as such will not depend on any fundamental constants except the overall energy scale, which scales linearly with the nucleon mass.  As this is a rather standard decay channel for nuclei of charge $4n+2$ \cite{kono}, and is needed to sidestep the otherwise 4th forbidden transition, we do not expect this channel to be modified upon variation of physical parameters.  Therefore, our expression for the total energy released is
\beq
E_{\text{tot}}=-w_0 m_e=E_b\left({}^{26}\text{Al}\right)-E_b\left({}^{26}\text{Mg}\right)-E_{\text{shift}}+m_n-m_p+E_2
\eeq

We need to relate the variables in this expression to more fundamental quantities, such as quark masses.  The primary dependence comes from the difference in the masses of the proton and neutron.  While the bulk of their rest mass is a result of the QCD froth they are comprised of, this will contribute to the two identically, leaving the remainder to arise from the mass difference of their constituent quarks and the electromagnetic forces between them.  The difference is found to be \cite{hn}
\beq
m_n-m_p=\Delta m -1.21 \frac{\alpha}{\alpha_0}\frac{m_p}{m_{p0}}\text{MeV}.
\eeq
Where $\Delta m=m_d-m_u=2.5\text{MeV}$ in our universe\footnote{A more accurate dependence would be found from lattice results for the electromagnetic splitting as in \cite{split}, which give $d(m_n-m_p)/d\Delta m=.93\pm.04$.  This is quantitatively very close to the more intuitive value of 1 that we use, and the choice does not affect any of our results.}.  While the quark masses will influence the total mass of the proton and neutron, this will be less than a percent level effect, and can be ignored.  When written in terms of the more fundamental parameters, this gives
\beq
w_0=2.31\frac{m_{e0}}{m_e}\left(\left(-3.01+6.13\frac{\alpha}{\alpha_0}\right)\frac{m_p}{m_{p0}}-2.12\frac{\Delta m}{\Delta m_0}\right)\text{MeV}
\eeq
Now that we have the expressions for the lifetime and energy of this decay, we can ask questions about how sensitive it is to the various underlying parameters, and how this constraint compares to the preexisting considerations found in the literature.

\section{Parameter Dependence}\label{pade}

\begin{centering}
\begin{figure*}[h]
\centering
\includegraphics[height=7.5cm]{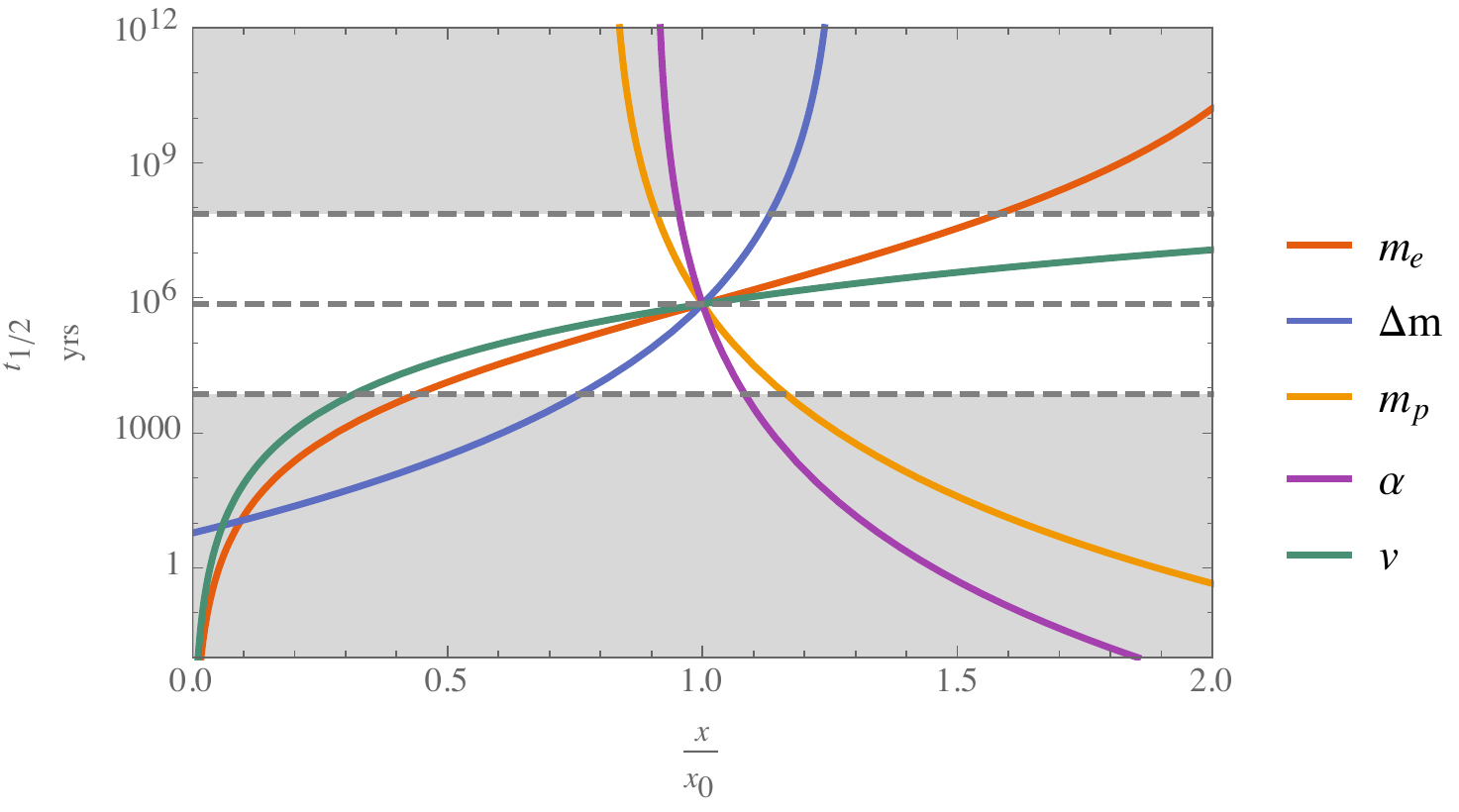}
\caption{Half life of aluminum-26 as a function of each parameter it depends on, normalized to their observed values.  }
\label{changes}
\end{figure*}
\end{centering}

In order to get a general idea of how sensitive the half life is to each of the parameters it depends on, we Taylor expand around their observed values, to find
\beq
t_{1/2}=717,000\text{yr}\left(1+7.29\frac{\delta m_e}{m_{e0}}+26.03\frac{\delta \Delta m}{\Delta m_{0}}-25.72\frac{\delta m_p}{m_{p0}}-75.32\frac{\delta\alpha}{\alpha_0}+4\frac{\delta v}{v_0}\right)
\eeq
In Fig. \ref{changes} we display how the half life varies over the full range.

From here the high sensitivity of the half life can be seen, as order one changes in all of these constants induce many orders of magnitude difference in the decay rate.  The half-life can be seen to depend inversely on the proton mass and fine structure constant, as increasing these will increase the binding energy.  However, the coefficients in this first order expansion are actually misleadingly small: for instance, if the proton mass is varied by $20\%$, the half life changes by 2-3 orders of magnitude.  The dependence on the electron mass and Higgs VEV are somewhat degenerate for smaller than observed values, a consequence of their both having simple power law scaling in this regime.  For larger values the half-life is least sensitive to the Higgs VEV, for exactly this reason as well. \\ 

\noindent{\bf Particle Masses:}
There has been much discussion on the anthropically allowed regions of quark and electron masses \cite{hn,closely,bbn,dd}.  It is generally found that order one deviations from the measured values would lead to a great variety of catastrophes that would preclude the existence of life as we know it.  This defines a catastrophic boundary \cite{catas} as the value of those parameters beyond which these changes take place.

The most immediate condition for the values of these masses is that the hydrogen atom must be stable to electron capture.  It is one of the most well known anthropic thresholds that if the neutron were just $3\%$ heavier, the electron orbitals in hydrogen would spontaneously collapse to form a neutron, preventing molecules such as water from existing.   In terms of quark and electron masses, this requirement is
\beq\label{hystab}
\Delta m-m_e-1.21\frac{\alpha}{\alpha_0}\frac{m_p}{m_{p0}}>0
\eeq
Some optimists hold out that life may still be possible in such a universe if it were made with deuterium, provided a scenario can be envisioned where it would be produced and distributed in abundance \cite{schell}.  If one permits this caution, then this requirement can be weakened to maintaining that some atom taken to be essential should be stable.  In \cite{hn} the blanket requirement that complex nuclei, with typical binding energies of order $8\text{MeV}$, should be stable, leading to a modification of eqn (\ref{hystab}).

A second requirement, more interesting for our purposes, was found in \cite{closely} by demanding the proton-proton chain to be exothermic, a requisite for stellar reactions to proceed.  This gives 
\beq\label{deut}
\Delta m+m_e-1.21\frac{\alpha}{\alpha_0}\frac{m_p}{m_{p0}}<2.2\frac{m_p}{m_{p0}}\text{MeV}.
\eeq
It has recently been suggested \cite{adamsdeut,barnesdeut} that this nuclear threshold may be circumvented through some other nuclear reaction that can be exothermic even if this one fails, such as $ppe\rightarrow D$.  Though stars may shine in this parameter regime, much more work needs to be done to determine if they are as conducive to life as they are in our universe (Do they have convective envelopes?  Will they flare often?  What is their UV flux, and long term evolution?  Can photosynthesis evolve around them?).  We therefore take the traditional bound as catastrophic for our analysis.

Now, we ask how our new requirement compares to these preexisting conditions.  The decay will not occur at all if $w_0<1$, which, when written in terms of the more fundamental parameters, gives
\beq
m_e+\Delta m < 3.68\frac{m_p}{m_{p0}}\left(1.97\frac{\alpha}{\alpha_0}-.97\right)\text{MeV}.
\eeq
The $\alpha$ dependent quantity in parentheses is normalized to 1 for the observed value of the fine structure constant.  It turns negative for values below $\alpha=1/278$, beyond which the decay is energetically impossible.

This limit is probably not the ultimate boundary for this mechanism- if the half life were much shorter, this isotope would decay far sooner than would be possible to disperse in the interstellar cloud.  If it were much longer, the cloud would disperse and planetary systems condense before it had a chance to leave its chiral imprint on the interstellar medium.  These considerations will be discussed in more detail below, but now we plot contour lines in the $m_e-\Delta m$ plane for several different decades of the half life, along with all the other criteria we have discussed until now.
\begin{centering}
\begin{figure*}[h]
\centering
\includegraphics[height=8.5cm]{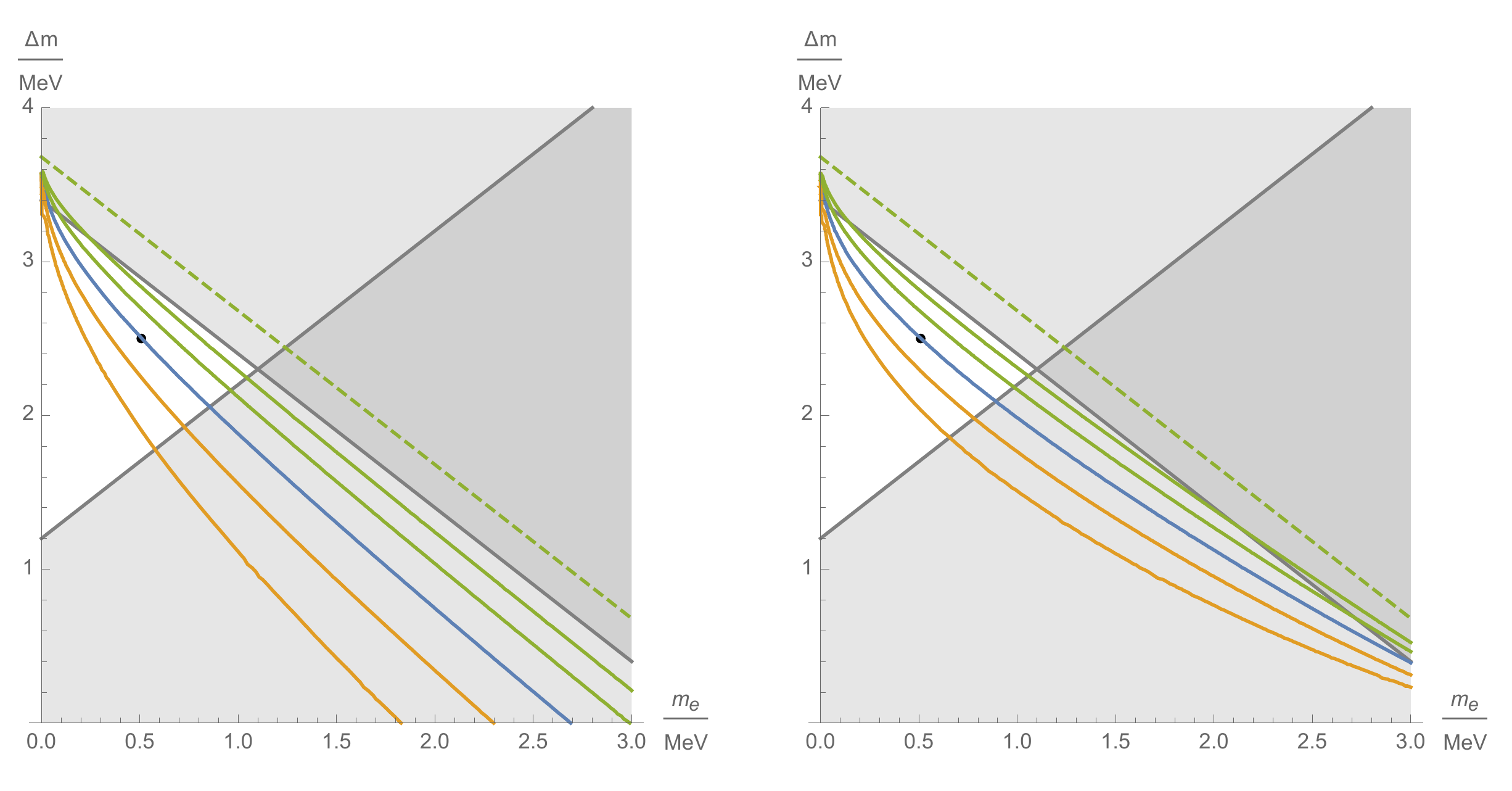}
\caption{Contours of constant half life of aluminum-26 in the $m_e-\Delta m$ plane.  The dot represents the observed values.  The orange curves below give $1/10$ and $1/100$ the half-life, and the green curves above give 10 and 100 larger than the observed value.  The dashed green line is the point at which the decay becomes energetically forbidden.  The shaded regions are the two other considerations mentioned in the text.  The two different plots are whether the Higgs VEV is taken to scale linearly with the quark masses (left), or be independent (right).}
\label{deltamme}
\end{figure*}
\end{centering}
\\ \\

We must also specify whether a change in the quark masses is concomitant with a change in the Higgs vacuum expectation value.  The masses being a product of the Yukawa couplings and this term $m_i=\lambda_i v$, our choice of which should be held fixed to affect an overall change will have different outcomes on the half life.  In Fig. \ref{deltamme} we present results for both choices, where the quark masses are linearly dependent on $v$, and where they are independent. 

Finally, we comment on the allowed range of the Higgs VEV, based off this criterion.  If the particle masses are taken to scale linearly with the Higgs VEV, then in order for the half life to be within a factor of 10 of its observed value, the range is restricted to be within $v/v_0\in\{.93,1.06\}$. Allowing for a factor of 100 gives $v/v_0\in\{.85,1.10\}$.  This can be compared to the bounds found in \cite{dd} based off the existence of hydrogen and heavy elements, where $v/v_0\in\{.39,1.64\}$ was found.  If the Yukawa couplings are allowed to compensate in order to hold the quark and lepton masses fixed, however, then the bounds become $v/v_0\in\{.56,1.78\}$ for a factor of 10 and $v/v_0\in\{.32,3.16\}$ for a factor of 100.  The crucial point here is that this mechanism lifts the previous degeneracies of these other bounds, which depend only on particle masses.

This sheds light on the idea of a weakless universe \cite{weakless}, which raises the question of whether the weak force is somehow necessary for life in our universe, given that its obvious effects are somewhat minimal.  There, the claim was that it is indeed possible to imagine a universe capable of producing life, even in its complete absence.  Criticism of this scenario already appeared in \cite{problems}, where it was pointed out that its lack of type II supernovae would prevent the creation of sufficient levels of oxygen, and even speculated that some connection to homochirality may be responsible as well.  Our analysis places this on a firm footing by linking the two by a concrete mechanism, which was recently bolstered by experiment.\\

\begin{centering}
\begin{figure*}[h]
\centering
\includegraphics[height=9.5cm]{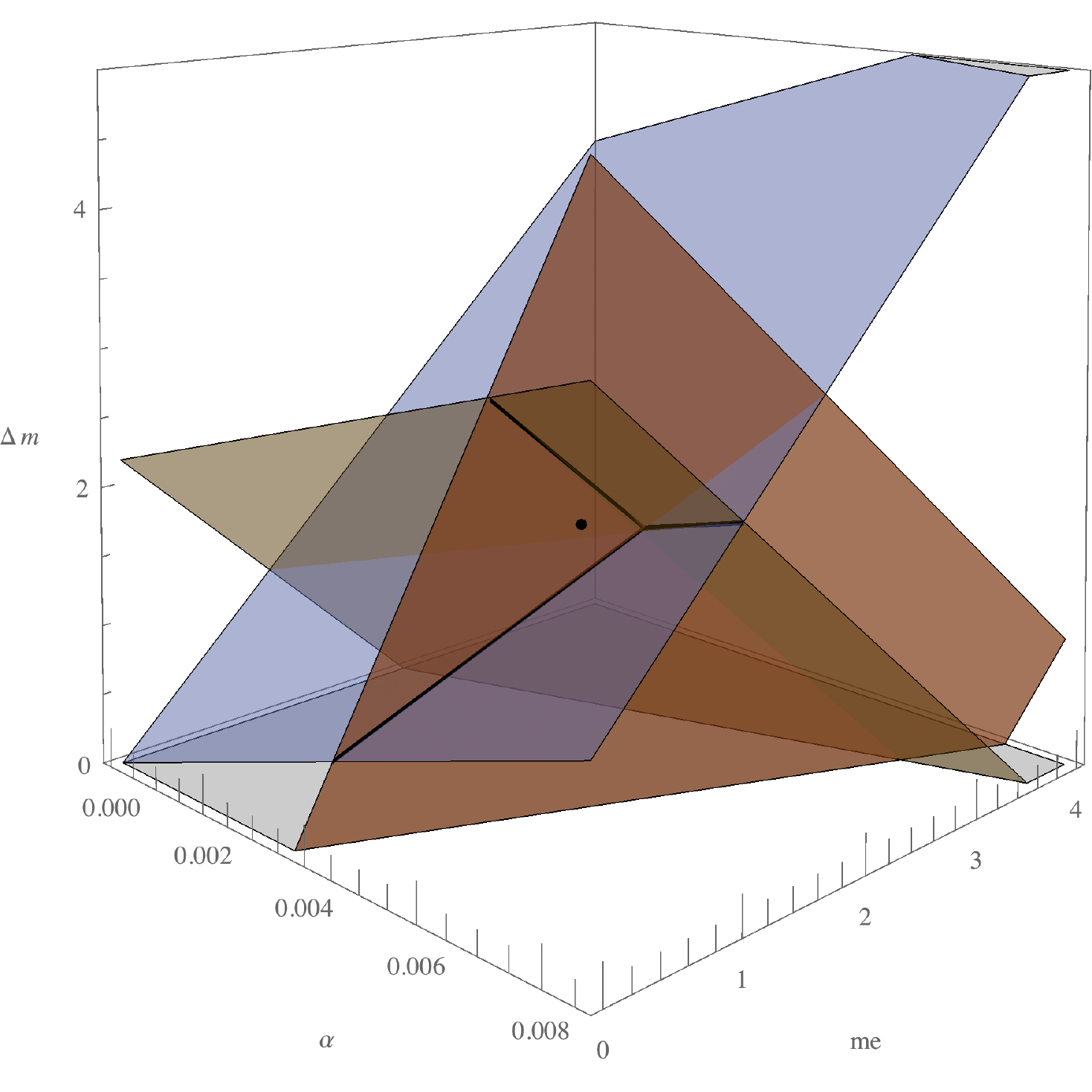}
\caption{Various thresholds in the $m_e-\alpha-\Delta m$ parameter space.  The yellow plane corresponds to the threshold for stellar fusion, the blue hydrogen stability, and the orange aluminum stability.  The dot is situated at the observed values of these parameters, and the allowed region is the one bounded by the three planes enveloping the dot.  The coincidence of the two thresholds is only a feature of the reduced $m_e-\Delta m$ plane because the observed value is close to a corner in the allowed space.}
\label{alphatoo}
\end{figure*}
\end{centering}

\noindent{\bf The Principle of Noncoincident Peril:}
One important feature to notice is that our new criterion is parallel to, and almost exactly coincident with, the requirement for stars.  This is difficult to interpret in the restricted terms of varying quark and electron mass.  We expect qualitative changes to the universe as we move about parameter space, but we do not expect multiple changes to occur at nearly the same transition.  We term this expectation the \emph{principle of noncoincident peril}.  This states that catastrophic boundaries in parameter space should not lead to simultaneous unrelated large changes in the world.  Furthermore, we take an observation of a violation of this principle as an indication that there must be some other parameter that we can also vary, that is able to shift at least one of these boundaries.  When viewed in this enlarged parameter space, the original coincidence is seen to be simply a corner in the allowed region that we happen to be situated close to, as illustrated in Fig. \ref{alphatoo}.  Note that in a multiverse context we expect to be situated close to the edges of the anthropically allowed region, if there is a preference for values that lie outside.  This has been dubbed the \emph{principle of living dangerously} \cite{justso}, and holds true for many fundamental constants \cite{matters}.

As a simple example illustrating this, consider the point in Fig. \ref{deltamme} where the two grey regions intersect, corresponding to $m_e=1.1\text{MeV}$ and $\Delta m=2.3\text{MeV}$.  If we were to think that the quark mass difference were held fixed, we would be puzzled by the fact that increasing the electron mass to $.9\text{MeV}$ would first render stars unstable, then shortly after, at $1.3\text{MeV}$, complex nuclei are unstable.  More puzzling still would be a hypothetical scenario where the observed difference between quark masses were $\Delta m=2.3\text{MeV}$, which would make these two transitions exactly coincident.  In the extended parameter space where $\Delta m$ is also allowed to vary, however, it is clear that there is only a special value for which these two conditions coincide.  An underlying preference for large electron mass would naturally push the observed value to be close to this corner \cite{bbn}.

The principle of noncoincident peril is a useful tool in the study of the multiverse, because rarely does a crucial physical process depend on only one parameter.  This complication usually introduces considerable uncertainties into attempts at inferring the existence of parameter tunings, subjecting them to provisos about the assumptions one makes about which parameters are variable.  A strong indication that a parameter must be taken as variable, then, is highly desired in order to remove these ambiguities.  Aside from assisting in these technical points, it also helps to answer an interesting question in itself: how many of the current fundamental constants are environmentally selected, and how many are set by the underlying theory of nature?  The principle of noncoincident peril provides a clue as to when a particular constant is necessarily variable.  

Let us apply our principle in the current setting.  Here, because the overlap of the aluminum and $pp$ fusion boundaries violates this principle, we can conclude that at least one of the other parameters in the expression for the half life, either $m_p$, $\alpha$, or $v$, must also be variable.  Notice that the dashed line $t_{1/2}=\infty$ corresponds to an `accumulation line' representing the asymptotic limit of the family of curves of constant half-life.  We take this line to be the actual threshold with which to compare to the stellar threshold.  This line corresponds to the condition $w_0=1$, which itself is a function of only $m_e$, $m_p$, $\Delta m$, and $\alpha$, i.e. it does not depend on the Higgs VEV.  Therefore, there are two additional parameters, the proton mass and the fine structure constant, that may be varied to change its $\Delta m$ intercept.  Additionally, the right hand side of (\ref{deut}), set by the deuteron binding energy, depends on both the mass of the proton and the strength of the strong force, $\alpha_s$.  In general, variations of these two parameters will in fact be correlated.  But, if we take $m_p$ to vary while $\alpha_s$ is held fixed (which can be achieved by varying the strange quark mass, for example), the effects of rescaling the relevant dimensionful quantity can be isolated from the other effect of varying the strength of the force.  Note that the dependence of both of these thresholds scale linearly with the proton mass.  Therefore, varying one would cause the other to vary proportionately, and including this parameter does not address the coincidence.  The only other choices, therefore, are the fine structure constants, $\alpha$ and $\alpha_s$.  For the remainder of this paper, we focus on the electromagnetic fine structure constant only, though we can only conclude that at least one of these must be variable.  But, if one is, it is quite plausible that the other is as well, so that our conclusion is that the fine structure constant(s) can vary in the multiverse.  

In Fig. \ref{alphatoo} we show this enlarged parameter space.  From here it becomes apparent that the coincidence of these two thresholds is not a generic feature in this full space, but rather only an apparent feature of the location of the observed values, which are situated in a corner of the allowed region.

First, we remark that there is no hidden correlation between these two thresholds.  This can be readily seen in the figure, since there are other values of the parameters for which the thresholds do not coincide.  This is in contrast to, for instance, the other threshold, that of the stability of complex elements.  In representing this plane we made a particular choice of the hydrogen stability threshold, but we could equally well have used another element, say carbon, oxygen, or nitrogen, to fit this demand.  All of these separate criteria fall on parallel planes that differ only by their overall height, which is set by the differences in binding energies between the appropriate neighboring isotopes, and are all of similar magnitude.  Therefore, even though all these separate conditions occur in neatly arranged succession, there is nothing suspicious about this fact, as they all represent slight variations on the same basic requirement.  In contrast, the enlarged parameter space demonstrates that there is no similar degeneracy between the requirement for stellar fusion to be operable and for aluminum-26 to have such a long half-life.

\begin{figure}
\centering
\begin{subfigure}
  \centering
  \includegraphics[height=8cm]{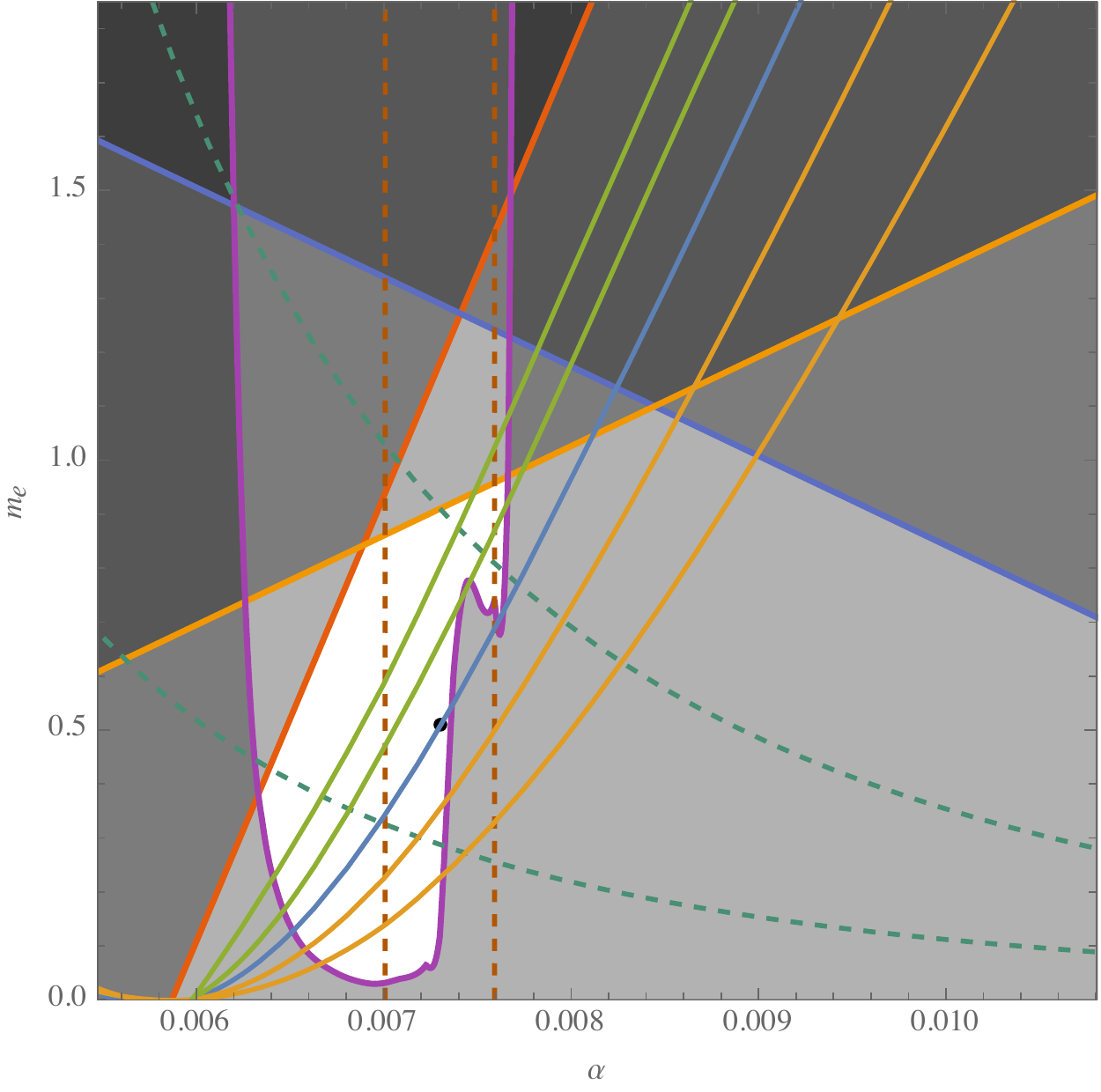}
  \label{fig:sub1}
\end{subfigure}%
\begin{subfigure}
  \centering
  \includegraphics[height=8cm]{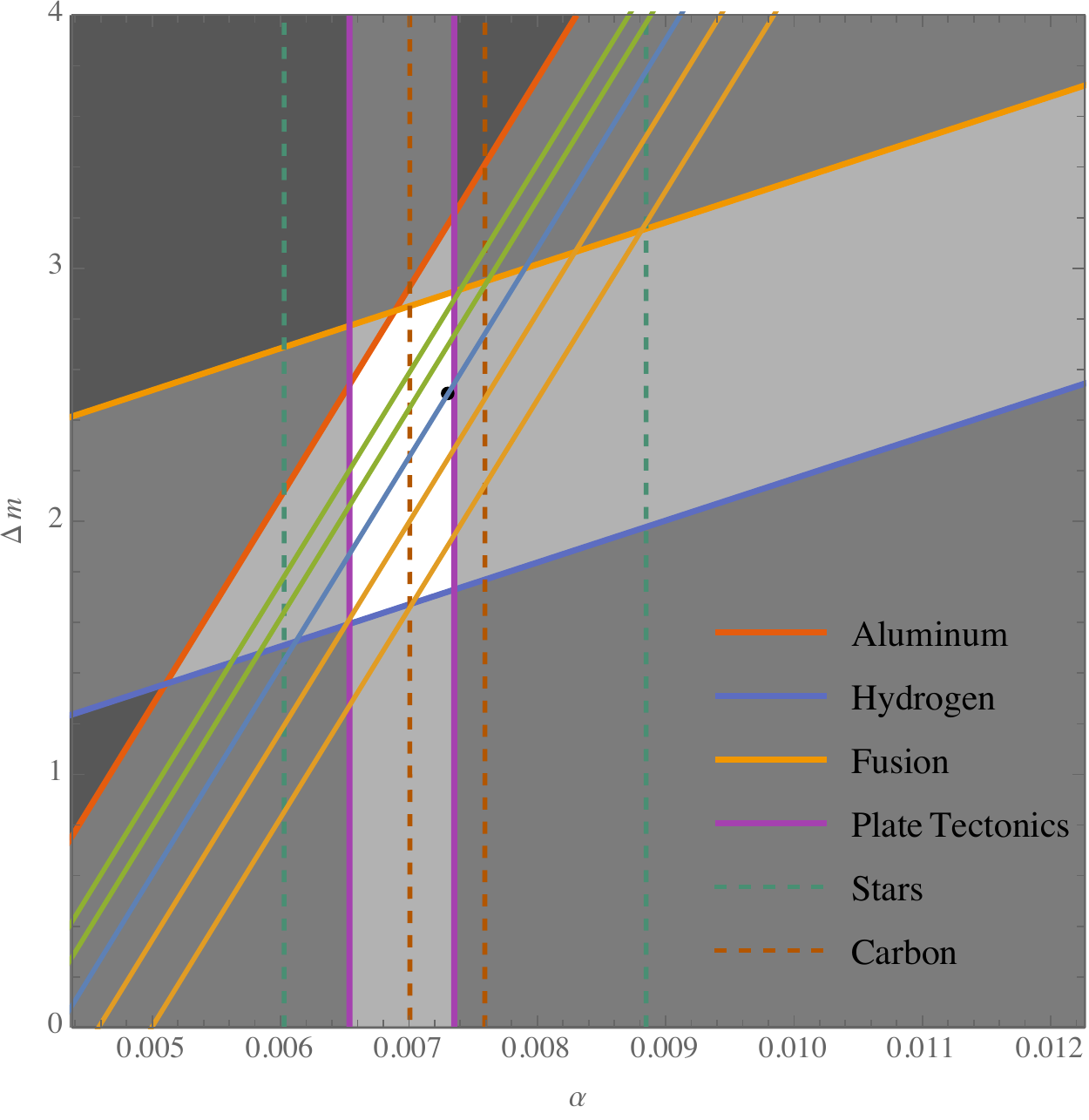}
  \label{fig:sub2}
\end{subfigure}
\caption{Allowed regions in the $\alpha-m_e$ and $\alpha-\Delta m$ planes, incorporating a variety of effects described in the text.}
\label{subplanes}
\end{figure}

Secondly, we may ask whether such a change in the fine structure constant is tolerable anthropically.   For this, we display several subplanes involving the fine structure constant, including other known boundaries, in Fig. \ref{subplanes}.


In addition to the boundaries already discussed, a few others from the literature are shown.  The requirement $\alpha^{12}m_e^4\sim m_p^6/M_{\text{Planck}}^2$, necessary for either the spectral temperature of our sun to be suitable for photosynthesis \cite{pl}, or the existence of both convective and radiative stars \cite{conrad}, is shown, labeled `Stars' (bands corresponding to equivalence within one order of magnitude).  Additionally, bounds found in \cite{ocs,ekllm} for the Hoyle resonance condition necessary for the production of carbon are shown, labeled `Carbon'.  Neither of these properties depend on the difference of quark masses, which is why they are vertical in the right plot.  Also included are the bounds for radioactive decay to be suitable for driving plate tectonics on terrestrial planets, found in \cite{sandy}.

One feature of these diagrams that is particularly noteworthy is the gross violation of the principle of noncoincident peril that seems to occur.  Increasing $m_e$ at constant $\alpha$ first encounters the fusion threshold, followed quickly by the aluminum threshold, and then the stellar condition.  This pile of coincidences is alleviated by recognizing that there is an additional dimensionless parameter on which the latter depends, namely the ratio of the proton mass to the Planck mass, which is capable of scaling the stellar curves up or down.  There is another coincidence between the narrow band needed for the Hoyle resonance and plate tectonics in the right plot, which may be taken to indicate that there are additional parameters which need to be variable.  From the work done in \cite{ocs,ekllm}, the Hoyle resonance is seen to depend also on a composite parameter representing the strength of the strong force, which is the ratio of the sum of the light quark masses to the scale of QCD breaking.  The efficacy of plate tectonics, which ultimately depends on the alpha decays of radioactive elements within the Earth, depends on the nuclear properties of alpha decay, as well as the habitable radius of terrestrial planets, as detailed in \cite{sandy}.\\

\noindent{\bf Setting:}
What is the range of half-lives for which the radioactive decay of aluminum is capable of imprinting chirality on the interstellar medium?  These unstable isotopes are created in short-lived massive stars, and then subsequently redistributed throughout the galactic environment by type II supernovae and Wolf-Rayet winds\cite{popsynth}.  In this process, a shock wave of the highest energy ejecta expands outward, with a relatively cooler interior of less energetic particles.  Even this interior region will be ionized, and would not be capable of carrying a chiral imprint.  After about $10^4$ years, the interior becomes cool enough for molecules to form\cite{chevalier}.  Taking the minimum requirement that the isotope be long lived enough to decay in a molecular environment, translates into the requirement that the half life must be $t_{1/2}>10^4$ yr, a factor of $70$ lower than the observed value.

The evolution of molecular clouds also provides an upper limit on the half life of the radioisotope: because erosive processes cause molecular clouds to continuously dissipate and reform on the timescale of 20 Myr \cite{turbmol}, any isotope that lived significantly longer than this would be unable to decay in time to leave a chiral imprint.  This is a factor of 28 times larger than the observed half life.  These exact numbers are not meant to represent absolute bounds, but are rather supposed to be indicative of the plausibly allowed variation.  A more detailed account of the galactic evolution would be necessary to understand the full extent to which the half life can be varied, which is beyond the scope of this analysis.\\


\noindent{\bf Alternate Decays:}
Finally, if the parameters were to be varied sufficiently so that the decay of aluminum-26 does not suffice, might some other beta decay do just as well?  To address this, we present Table \ref{tableone} of all elements that undergo beta decay with half lives greater than a year and less than a billion, compiled from \cite{nuc}.
\begin{table}[tb]
\vskip.4cm
\begin{center}
\begin{tabular}{|c|c|c|c|}
\hline Isotope & $t_{1/2}$(years) & decay & produced ($M_{\odot}$) \\
\hline
{\color{red}${}^{3}$H} & {\color{Goldenrod}12} & {\color{red}$\beta^-$} & {\color{red}$2.8\times10^{-19}$}\\
{\color{red}${}^{10}$Be} & $1.5\times10^6$ & {\color{red}$\beta^-$} & {\color{red}$3.8\times10^{-24}$}\\
{\color{red}${}^{14}$C} & $5700$ & {\color{red}$\beta^-$} & $6.7\times10^{-9}$\\
{\color{Goldenrod}${}^{22}$Na} & {\color{Goldenrod}$2.6$} & $\beta^+$ & $9.8\times10^{-8}$\\
{\color{green}${}^{26}$Al} & {\color{green}$717,000$} & {\color{green}$\beta^+$/ec} & {\color{green}$1.8\times10^{-6}$}\\
{\color{red}${}^{32}$Si} & {\color{Goldenrod}$153$} & {\color{red}$\beta^-$} & $1.3\times10^{-8}$\\
{\color{red}${}^{36}$Cl} & $3.0\times10^5$ & {\color{red}$\beta^-$} & $5.0\times10^{-7}$\\
{\color{red}${}^{39}$Ar} & {\color{Goldenrod}$269$} & {\color{red}$\beta^-$} & $3.5\times10^{-9}$\\
{\color{Goldenrod}${}^{40}$K} & {\color{Goldenrod}$1.2\times10^9$} & $\beta^-/\beta^+$/ec & $1.0\times10^{-8}$\\
{\color{red}${}^{42}$Ar} & {\color{Goldenrod}$32.9$} & {\color{red}$\beta^-$} & {\color{red}$3.1\times10^{-13}$}\\
{\color{red}${}^{59}$Ni} & $7.6\times10^4$ & {\color{red}ec} & $1.8\times10^{-6}$\\
{\color{red}${}^{60}$Co} & {\color{Goldenrod}$5.3$} & {\color{red}$\beta^-$} & {\color{red}$6.3\times10^{-14}$}\\
{\color{red}${}^{60}$Fe} & $2.6\times10^6$ & {\color{red}$\beta^-$} & {\color{red}$1.8\times10^{-17}$}\\
{\color{red}${}^{63}$Ni} & {\color{Goldenrod}$101$} & {\color{red}$\beta^-$} & {\color{red}$5.5\times10^{-16}$}\\
\hline
\end{tabular}
\end{center}
\caption{Complete list of candidate isotopes potentially capable of imprinting chirality on a molecular cloud.  Entries in red do not meet the criteria necessary to operate.  Those in yellow are inoperational in our universe, but may potentially be suitable with a large enough change in the physical constants.  Numbers in the last column are taken from \cite{ccej} for a 13 solar mass supernova.}
\label{tableone}
\end{table}
Though there are fourteen elements on the list, all of them are less suitable for imprinting a chiral signature in the interstellar medium.  Nearly half of them are not produced in the same abundance by type II supernovae, making their imprints orders of magnitude weaker.  Over half of them have half lives that are more than three orders of magnitude away from the observed half-life of aluminum.  This is a less strict criterion, as it is possible to affect a change of this magnitude by relatively mild variations of the fundamental parameters.

However, the type of decay is the most serious discriminator for an isotope's relevance.  As mentioned before, if the dominant channel is electron capture, then a nonchiral photon will be produced, and the mechanism is completely absent.  It is worth noting that, depending on the case, the branching ratio between positron emission and electron capture may increase appreciably if the energy increases (concomitantly decreasing the half-life), making ${}^{40}$K in particular a candidate that is not ruled out by this criterion.  

Perhaps more surprising is the observation that electron decays are less suitable than positrons.  Though both decays imprint the same sign of chirality, the kinematics of how the two interact with interstellar molecules is very different.  This is due to the fact that while positrons scatter off electrons by exchange glancing blows and continue on relatively unimpeded paths, electrons can also backscatter, where they reverse direction almost completely (in the center of mass frame- in the rest frame, this corresponds to a $90^\circ$ split).  While any scattering angle will preserve helicity at high energies, the crucial point is that the results of \cite{dg} show a peak asymmetry below 10 eV.  In this nonrelativistic regime, the corruption of initial helicity is common for wide scattering angles, and so electrons lose their initial chirality much faster than positrons.  We find that the closest possible alternative isotopes, ${}^{40}$K and ${}^{22}$Na, are 3 and 5 orders of magnitude away from fulfilling our requirement, respectively.

\section{Discussion}\label{codi}
Though the origin of biomolecular homochirality, an aspect of chemistry thought to be crucial for life, remains unknown, the radioactive decay of aluminum-26 in the interstellar medium is a leading candidate for explanation.  If true, this is reliant on a seemingly fine-tuned binding energy, that makes this decay unusually long lived.  Moderate variations of various physical parameters can drastically change this, rendering the entire mechanism inoperational.  Additionally, for the threshold of this occurrence to not inadvertently coincide with the stability of deuterium, in contradiction with the principle of noncoincident peril, the fine structure constant must be taken to be variable.  

Asking anthropic questions at this level of detail will become more and more feasible as we learn more about life, its origins, and its possible environments.  It has the potential for clarifying the border between which aspects of the observed laws of physics are set by a selection effect predicated on the existence of observers, and which can have a more fundamental explanation.  Mapping the dependence of the requirements for complex life is a promising avenue for differentiating between the two.  As it stands now, there are several dozen low energy parameters that have yet to be explained by a more fundamental theory, and considerably less constraints coming from anthropic considerations.  It is only through efforts to uncover as many fine tunings as possible that we will be able to find out how the number of constraints compares to the number of fundamental constants.  If, after exhaustive search, we find less constraints than parameters, the universe will be known to be rather flexible in its properties.  If exactly equal, we can infer that the parameters are rigidly set.  Uncovering a greater number of requirements than tunable parameters will yield a scenario that would be very challenging to interpret.

One encouraging aspect of this particular investigation is that while it is speculative, it will be testable in the coming decades.  Future observations of the presence of chirality in interstellar clouds, other locales within our solar system, or remote sensing of other star systems will be able to support or rule out this hypothesis completely.  This demonstrates that further information regarding the multiverse can be gleaned from future experiments in astronomy, planetary science, and origin of life research.

So did the universe take such care to fiddle with its parameters at such a level as to help our particular biochemistry get started?  Why not?  After all, if the main success of anthropic reasoning is that the universe tunes the cosmological constant by 120 orders of magnitude away from its natural value, surely the possibility that it tune by another order of magnitude or so to provide a more fecund chemistry is not so outlandish.

The Vester-Ulbricht hypothesis, while still unproven, has been used here to illustrate a general programme: one that draws upon cutting-edge developments in the science of habitability across a broad range of disciplines, in order to continually refine our notions of the multiverse hypothesis.  The issue of possible alternative universes should rightfully be regarded as one of the most challenging scientific questions of our age, if not ever, and to have a hope of shedding light on this elusive question, we must be willing to harness knowledge from essentially all branches of science.  We continue to learn much about life, its requirements to survive and thrive, and there is every indication that we will continue to learn even more in the coming decades.  It would be imprudent to not ask what implications our newfound knowledge has for our place in the universe, and potentially beyond.\\

{\bf \noindent Acknowledgements:}
I would like to thank David Catling, Jaume Garriga, Brett McGuire, Karin Oberg, Ken Olum, and Patrick Slane for useful discussions.


\smallskip

\begin{thebibliography}{99}

\bibitem{fitness}
L. J. Henderson, \emph{The Fitness of the Environment}. Macmillan, 1913.

\bibitem{water}
R. A. King, A. Siddiq, W. D. Allen and H. F. Schaefer, ``Chemistry as a function of the
fine-structure constant and the electron-proton mass ratio." Phys. Rev. A 81, 042523
(2010) doi:10.1103/PhysRevA.81.042523.

\bibitem{hard}
S. Walker, and P. Davies, ``The `Hard Problem' of Life", 
In S.I. Walker, P.C.W. Davies and G.F.R. Ellis (eds): \emph{From Matter to Life: Information and Causality}. Cambridge University Press, 2017
arXiv:1606.07184.

\bibitem{VU}
T. L. V. Ulbricht, and F. Vester, ``Attempts to induce optical activity with polarized $\beta$-radiation." Tetrahedron {\bf 18.5} 629 (1962).
doi:http://dx.doi.org/10.1016/S0040-4020(01)92714-0.

\bibitem{cline}
D.B. Cline,
``Possible Physical Mechanisms in the Galaxy to Cause Homochiral Biomaterials for Life,"
Symmetry {\bf 2.3} 1450 (2010).  
doi:10.3390/sym2031450.

\bibitem{podlech}
J. Podlech,
"Origin of organic molecules and biomolecular homochirality." 
Cellular and Molecular Life Sciences {\bf 58} 44 (2001).
doi:10.1007/PL00000777.

\bibitem{gk}
V.I. Goldanskii and V. V. Kuzmin,
``Spontaneous breaking of mirror symmetry in nature and the origin of life." 
Physics-Uspekhi {\bf 32} 1 (1989).
doi:10.1070/PU1989v032n01ABEH002674.

\bibitem{bonner}
W.A. Bonner,
``The origin and amplification of biomolecular chirality." 
Origins of Life and Evolution of Biospheres {\bf 21} 59 (1991).


\bibitem{keszthelyi}
L. Keszthelyi,
"Origin of the homochirality of biomolecules." 
Quart. Rev. Biophys. {\bf 28} 473 (1995).

\bibitem{morozov}
L. Morozov, 
``Mirror symmetry breaking in biochemical evolution." 
Origins of Life and Evolution of Biospheres {\bf 9.3} 187 (1979)
doi:10.1007/BF00932495.

\bibitem{agk}
V. A. Avetisov, V. V. Kuz'min, and V. I. Goldanskii,
``Handedness, origin of life and evolution." Physics Today 44.7 (1991): 33-41.
doi:10.1063/1.881264.

\bibitem{jbg}
F. Jafarpour, T. Biancalani, and N. Goldenfeld,
``Noise-induced mechanism for biological homochirality of early life self-replicators." 
Phys. Rev. Lett. {\bf 115.15} 158101 (2015)
doi:10.1103/PhysRevLett.115.158101
arXiv:1507.00044


\bibitem{pdiff}
S. Mason, and G. Tranter,
``The parity-violating energy difference between enantiomeric molecules." 
Mol. Phys. {\bf 53} 1091 (1984)
doi:10.1080/00268978400102881.

\bibitem{lwc} 
  Y.~Liu, H.~Wang and D.~Cline,
  ``Simulation of a weak interaction induced chiral transition in a prebiotic medium,''
  AIP Conf.\ Proc.\  {\bf 300}, 499 (1994).
  doi:10.1063/1.45443.

\bibitem{nonlin}
C. Girard, and H. B. Kagan,
``Nonlinear effects in asymmetric synthesis and stereoselective reactions: ten years of investigation." 
Angewandte Chemie Int. Ed. {\bf 37.21}, 2922 (1998).
doi:10.1002/(SICI)1521-3773.

\bibitem{murch}
J.R. Cronin and S. Pizzarello,
``Enantiomeric excesses in meteoritic amino acids." 
Science {\bf 275} 951 (1997).
doi:10.1126/science.275.5302.951


\bibitem{cs}
C. Chyba and C. Sagan,
``Endogenous production, exogenous delivery and impact-shock synthesis of organic molecules: an inventory for the origins of life,"
Nature {\bf 355} 125 (1992).
doi:10.1038/355125a0.


\bibitem{shocks}
Martins, Zita, et al. 
``Shock synthesis of amino acids from impacting cometary and icy planet surface analogues," 
Nature Geosci {\bf 6.12} 1045 (2013).
doi:10.1038/ngeo1930.

\bibitem{stardust}
S. Sandford et al,
``Organics captured from comet 81P/Wild 2 by the Stardust spacecraft.'' 
Science {\bf 314} 1720 (2006).
doi:10.1126/science.1135841.

\bibitem{cosac}
F. Goesmann et al,
``Organic compounds on comet 67P/Churyumov-Gerasimenko revealed by COSAC mass spectrometry,"
Science {\bf 349} (2015).
doi:10.1126/science.aab0689.


\bibitem{candc}
B. Cohen, and R. Coker, 
``Modeling of liquid water on CM meteorite parent bodies and implications for amino acid racemization." 
Icarus {\bf 145.2} 369 (2000)
doi:10.1006/icar.1999.6329.  


\bibitem{glue}
B.A. McGuire et al, 
``Discovery of the interstellar chiral molecule propylene oxide (CH3CHCH2O),"
Science {\bf 352} 1449 (2016).
doi:10.1126/science.aae0328
[arXiv:1606.07483v1 [astro-ph.GA]]


\bibitem{cplexpt}
J. Bailey et al,
``Circular Polarization in Star-Formation Regions: Implications for Biomolecular Homochirality,"
Science {\bf 281} 672 (1998).
doi:10.1126/science.281.5377.672.

\bibitem{CPLtheory}
P. W. Lucas, \emph{et al.}, ``UV circular polarisation in star formation regions: the origin of homochirality?." Origins of Life and Evolution of Biospheres {\bf 35} 29 (2005)
doi:10.1007/s11084-005-7770-6.

\bibitem{dg}
J.M. Dreiling and T. J. Gay, 
"Chirally sensitive electron-induced molecular breakup and the Vester-Ulbricht hypothesis,"
Phys. Rev. Lett. {\bf 113} 118103 (2014)
doi:10.1103/PhysRevLett.113.118103.

\bibitem{meiring}
W.J.Meiring,
``Nuclear Beta-decay and the origin of biological chirality,"
Nature {\bf 329} 712 (1987)
doi:10.1038/329712a0.

\bibitem{patty}
C.H.L. Patty et al,
``Circular spectropolarimetric sensing of chiral photosystems in decaying leaves,"
J. Quant. Spec. and Rad. Trans. {\bf 189} 303 (2017).
doi:10.1016/j.jqsrt.2016.12.023
[arXiv:1701.01297 [q-bio.BM]]

\bibitem{CEA}
``Table de Radiocucl\'eides,"
CEA ISBN 2 7272 0200 8.


\bibitem{laubitz}
M.J. Laubitz,
``The Beta Decay of the Ground State of 26Al,"
Proc. of the Phys. Soc. {\bf 68} 1033 (1955).
doi:10.1088/0370-1298/68/11/311.


\bibitem{krane}
K. Krane,
{\it Introductory Nuclear Physics, 3rd Ed.},
Wiley, Singapore,1988,
ISBN 978-0-471-80553-3.



\bibitem{br} 
  H.~Brysk and M.~E.~Rose,
  ``Theoretical Results on Orbital Capture,''
  Rev.\ Mod.\ Phys.\  {\bf 30}, 1169 (1958).
  doi:10.1103/RevModPhys.30.1169

  \bibitem{ku} 
  E.~J.~Konopinski and G.~E.~Uhlenbeck,
  ``On the Fermi Theory of beta-Radioactivity,''
  Phys.\ Rev.\  {\bf 48}, 7 (1935).
  doi:10.1103/PhysRev.48.7.


\bibitem{kono} 
  E.~J.~Konopinski,
  ``Beta-Decay,''
  Rev.\ Mod.\ Phys.\  {\bf 15}, 209 (1943).
  doi:10.1103/RevModPhys.15.209.


\bibitem{hn} 
  L.~J.~Hall and Y.~Nomura,
  ``Evidence for the Multiverse in the Standard Model and Beyond,''
  Phys.\ Rev.\ D {\bf 78}, 035001 (2008)
  doi:10.1103/PhysRevD.78.035001
  [arXiv:0712.2454 [hep-ph]].

\bibitem{split} 
  A.~W.~Thomas, X.~G.~Wang and R.~D.~Young,
  ``Electromagnetic Contribution to the Proton-Neutron Mass Splitting,''
  Phys.\ Rev.\ C {\bf 91}, no. 1, 015209 (2015)
  doi:10.1103/PhysRevC.91.015209
  [arXiv:1406.4579 [nucl-th]].

\bibitem{closely} 
  C.~J.~Hogan,
  ``Quarks, electrons, and atoms in closely related universes,''
  In \emph{Carr, Bernard (ed.): Universe or multiverse?} 221-230
  [astro-ph/0407086].

\bibitem{bbn} 
  L.~J.~Hall, D.~Pinner and J.~T.~Ruderman,
  ``The Weak Scale from BBN,''
  JHEP {\bf 1412}, 134 (2014)
  doi:10.1007/JHEP12(2014)134
  [arXiv:1409.0551 [hep-ph]].

\bibitem{dd} 
  T.~Damour and J.~F.~Donoghue,
  ``Constraints on the variability of quark masses from nuclear binding,''
  Phys.\ Rev.\ D {\bf 78}, 014014 (2008)
  doi:10.1103/PhysRevD.78.014014
  [arXiv:0712.2968 [hep-ph]].

\bibitem{catas} 
  R.~Bousso, L.~J.~Hall and Y.~Nomura,
  ``Multiverse Understanding of Cosmological Coincidences,''
  Phys.\ Rev.\ D {\bf 80}, 063510 (2009)
  doi:10.1103/PhysRevD.80.063510
  [arXiv:0902.2263 [hep-th]].

\bibitem{schell} 
  A.~N.~Schellekens,
  ``Life at the Interface of Particle Physics and String Theory,''
  Rev.\ Mod.\ Phys.\  {\bf 85}, no. 4, 1491 (2013)
  doi:10.1103/RevModPhys.85.1491
  [arXiv:1306.5083 [hep-ph]].

\bibitem{adamsdeut} 
  F.~C.~Adams and E.~Grohs,
  ``On the Habitability of Universes without Stable Deuterium,''
  Astropart. Phys. {\bf 91} 90 (2017)
  doi:10.1016/j.astropartphys.2017.03.009 
  arXiv:1612.04741 [astro-ph.CO].
  
\bibitem{barnesdeut} 
  L.~A.~Barnes and G.~F.~Lewis,
  ``Producing the Deuteron in Stars: Anthropic Limits on Fundamental Constants,''
  arXiv:1703.07161 [astro-ph.CO].

\bibitem{weakless} 
  R.~Harnik, G.~D.~Kribs and G.~Perez,
  ``A Universe without weak interactions,''
  Phys.\ Rev.\ D {\bf 74}, 035006 (2006)
  doi:10.1103/PhysRevD.74.035006
  [hep-ph/0604027].

\bibitem{problems} 
  L.~Clavelli and R.~E.~White, III,
  ``Problems in a weakless universe,''
  hep-ph/0609050.

\bibitem{justso} 
  C.~J.~Hogan,
  ``Why the universe is just so,''
  Rev.\ Mod.\ Phys.\  {\bf 72}, 1149 (2000).
  doi:10.1103/RevModPhys.72.1149
  [astro-ph/9909295].

\bibitem{matters} 
  M.~Tegmark, A.~Aguirre, M.~Rees and F.~Wilczek,
  ``Dimensionless constants, cosmology and other dark matters,''
  Phys.\ Rev.\ D {\bf 73}, 023505 (2006)
  doi:10.1103/PhysRevD.73.023505
  [astro-ph/0511774].

\bibitem{pl} 
  W.~H.~Press and A.~P.~Lightman,
  ``Dependence Of Macrophysical Phenomena On The Values Of The Fundamental Constants,''
  Submitted to: Phil. Trans. Roy. Soc. (Lond) A.

\bibitem{conrad}
  B.~Carter,
  ``Large number coincidences and the anthropic principle in cosmology,''
  IAU Symp.\  {\bf 63}, 291 (1974).

\bibitem{ocs} 
  H.~Oberhummer, A.~Csoto and H.~Schlattl,
  ``Stellar production rates of carbon and its abundance in the universe,''
  Science {\bf 289}, 88 (2000)
  doi:10.1126/science.289.5476.88
  [astro-ph/0007178].

  \bibitem{ekllm} 
  E.~Epelbaum {\it et al.},
  ``Dependence of the triple-alpha process on the fundamental constants of nature,''
  Eur.\ Phys.\ J.\ A {\bf 49}, 82 (2013)
  doi:10.1140/epja/i2013-13082-y
  [arXiv:1303.4856 [nucl-th]].  

\bibitem{sandy} 
  M.~Sandora,
  ``The Fine Structure Constant and Habitable Planets,''
  JCAP {\bf 1608}, no. 08, 048 (2016)
  doi:10.1088/1475-7516/2016/08/048
  [arXiv:1604.03151 [astro-ph.CO]].


\bibitem{popsynth}  
R. Voss, et al, 
``Population synthesis models for 26 Al production in starforming regions." 
New Astro. Rev. {\bf 52.7}, 436 (2008)
doi:10.1016/j.newar.2008.06.022.
  
\bibitem{chevalier}  
R. Chevalier, 
``Supernova remnants in molecular clouds." 
The Astrophys. J. {\bf 511.2} 798 (1999)
doi:10.1086/306710
[astro-ph/9805315].

\bibitem{turbmol}
R. B. Larson,
``Turbulence and star formation in molecular clouds." 
Monthly Not. of the Royal Astro. Soc. {\bf 194.4} 809 (1981)
doi:10.1093/mnras/194.4.809.

\bibitem{nuc}
"Live Chart of Nuclides",
https://www-nds.iaea.org/relnsd/vcharthtml/VChartHTML.html
Web. Feb 21, 2017.

\bibitem{ccej}
F. Thielemann, K. Nomoto, and M. Hashimoto,
``Core-collapse supernovae and their ejecta." 
The Astrophys. J. {\bf 460} 408 (1996)
doi:10.1086/176980.



\end{thebibliography}
\end{document}